\RequirePackage{fix-cm}
\documentclass[smallextended]{svjour3}       
\smartqed  
\usepackage{graphicx}
\journalname{Experimental Astronomy}
\begin{document}

\title{Digital Complex Correlator for a C-band Polarimetry survey}



\author{Miguel Bergano  \and
        Francisco Fernandes \and 
        Lu\'{\i}s Cupido \and
        Domingos Barbosa \and
        Rui Fonseca \and
        Ivan Ferreira \and
        Bruce Grossan \and
        George Smoot
}

\authorrunning{Bergano et al.} 

\institute{Miguel Bergano \and Domingos Barbosa \and Rui Fonseca \at
             Radio Astronomy Group,  Instituto de Telecomunica\c c\~oes, Campus Universit\'ario de Santiago, 3810-193 Aveiro, Portugal \\
              Tel.: +351-234-377900\\
              Fax: +351-234-377901\\
              \email{jbergano@av.it.pt}           
           \and
           Francisco Fernandes \at
              Currently at Outsoft Ltd., Ed. Colombo I, Av. Dr. M\'ario Sacramento, 177 1ºN, 3810-106 Aveiro, Portugal
              \and
           Lu\'{\i}s Cupido \at
              Instituto de Plasmas  e Fus\~ao Nuclear, Instituto Superior T\'ecnico, Av. Rovisco Pais, 1049-001 Lisboa, Portugal\\
              \email{cupido@ua.pt}
              \and
           Rui Fonseca \at 
           Universidade de Aveiro- Campus Universit\'ario de Santiago, 3810-193 Aveiro, Portugal 
           \and
           Ivan Ferreira \at
           Laborat\'orio de Plasmas, Instituto de F\'{\i}sica, University of Bras\'{\i}lia, Caixa Postal 04455, 70919-970 Bras\'{\i}lia - DF, Brasil \\
           and\\
           Instituto Nacional de pesquisas Espaciais\\
           Av. dos Astronautas 1.758 - Jd. Granja - CEP 12227-010, S\~ao Jos\'e dos Campos, SP - Brasil
           \and
           Bruce Grossan \and George Smoot \at
           Lawrence Berkeley National Laboratory , University of California, 1 Cyclotron Road, Bldg. 50, MS 205, Berkeley, CA 94720, USA\\
           Tel.:+15104865237\\
           Fax: +15104867149\\
           \email{Bruce\_Grossan@lbl.gov}
}

\date{Received: date / Accepted: date}

\maketitle

\begin{abstract}
The international Galactic Emission Mapping project aims to map and characterize the polarization field of the Milky Way. In Portugal it will cartograph the C-band sky polarized emission of the Northern Hemisphere and provide templates for map calibration and foreground control of microwave space probes like ESA Planck Surveyor mission. The receiver system is equipped with a novel receiver with a full digital back-end using an Altera Field Programmable Gate Array, having a very favorable cost/performance relation. This new digital backend comprises a base-band complex cross-correlator outputting the four Stokes parameters of the incoming polarized radiation. In this document we describe the design and implementation of the complex correlator using COTS components and a processing FPGA, detailing the method applied in the several algorithm stages and suitable for large sky area surveys. 
\end{abstract}

\section{Introduction}
\label{sec:1}
The Galactic Emission Mapping project (GEM) [6, 7] started as a tool to map the foreground synchrotron radiation originated within our galaxy in radio spectral bands and its contamination on the Cosmic Microwave Background (CMB).  In recent years it has been evolving towards a radio polarimetry survey experiment targeting the mapping of large-sky areas and provide information on the polarization field of the Milky Way with a ongoing 5GHz South Hemisphere already on the way (Wuensche 2009). This decade we assisted to the onset and detailed mapping of the CMB polarization, started by DASI experiment first detections in 2002 (Kovac et al., 2002; Leitch et al. 2002), the first information by the WMAP satellite on large scale polarized signal (Jarosik et al. 2009; Dunkley et al. 2010) and the soon oncoming first all-sky templates of CMB polarization from the Planck Surveyor space mission covering a wide range of frequencies (30-850GHz) (Efstathiou et al, Planck Blue Book).  To improve information on the polarized galactic foreground emission, GEM project led to the development of stable receivers capable of detecting the polarized synchrotron continuum at lower frequencies. The necessity for high sensitivity implying a relatively large bandwidth and concerns about Faraday rotation depolarization motivated the focus of observation centered in the 5GHz band (C-band). At this frequency, Faraday rotation is negligible enough and so templates provide a good extrapolation of the polarization characteristics of synchrotron emission to much higher microwave frequency bands and are of good use to synchrotron studies in general for the new generation of radio astronomy projects. More recently, in Portugal, for the northern part of the project, we adopted a 200MHz receiver bandwidth, centered at 4.9GHz, and free from Radio frequency Interference. In later project stages we will expand the bandwidth to 400GHz, if radio spectral conditions at the observing site (Fonseca et al., 2006) maintain as very good, implying a necessary adaptation of the correlator developed along the principles here outlined.

\section{The C-band receiver }
\label{sec:1}
The GEM project started originally in South Hemisphere – Brazil (Torres et al.,1996; Tello et a. 2007), having a recent counterpart addition in the North Hemisphere, after the installation of a modified telecom antenna in central Portugal in a radio quiet site (Barbosa et al. 2010, in preparation). Together, they will map more than 85\% of the sky. The final application of these joint surveys will be the separation of the several component emissions from the CMBR and the Milky Way by experiments like the satellite Planck Surveyor. For the northern part of  sky and motivated by the needs of receiver stability  we took advantage of the good performance of nowadays COTS technologies and so adopted a superheterodyne solution for the receiver with a digital backend, with higher immunity to offsets introduced by the custom analogue electronics. This super heterodyne receiver will be located in the focus of a 9-meter Cassegrain antenna and aims to achieve a sensibility of 300 µK after 6 months of operation with 50\% efficiency. The telescope has a final beam width of about 45 arcmin and the adopted GEM survey strategy  means it is  capable of azimuthal rotation at constant elevation (60 deg), about 0.3 rpm of constant speed, so each sky pixel is observed about 0.5 seconds (Grossan et al., in preparation). The azimuthal rotation if motivated by the needs of minimization of atmosphere and receiver noise fluctuations within a circular scan. Thus, the combination of bandwidth and pixel integration time will be key parameters for the correlator algorithm to be implemented. The super heterodyne receiver chain schematics is shown in Figure 1 and basically its gain properties are such it can amplify the sky signal into the detection range provided by the ADCs dynamical range.
\begin{figure}
 \includegraphics[width=1\textwidth]{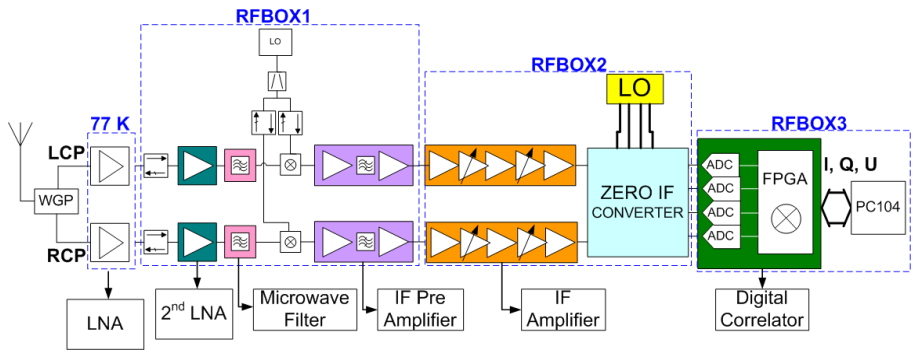}
\caption{Simple schematics of the receiver chain. Each box is temperature controled.}
\label{fig:1}       
\end{figure}

The incoming wave collected by the antenna is split in the two circular polarization hands (RCP – Right Circular Polarization and LCP- Left Circular Polarization) by the Wave Guide Polarizer (WGP). These signals are immediately amplified by cryocooled InP based Low Noise Amplifier (LNA) from The Low Noise Factory, Chalmers (model LNF-LNC4\_8A), followed by a second LNA working at room temperature (Bergano et al. 2007). The inclusion of this second RF LNA decreases the noise contribution of the following stages, the filter, the mixer and subsequent IF chain. The RF signal is down converted to 600 MHz where it is amplified and filtered by an IF chain with 200MHz of bandwidth consisting of a low noise IF amplifier followed by an high gain amplifier in which the gain can be set. It is finally converted to Zero IF (baseband) in the form of in phase and quadrature components and these signals are digitized by four Analog to Digital Converters (ADC) at 200Ms/s sampling rate, feeding a Field Programmable Gate Array (FPGA) working at 100MHz where the cross correlation processing happens. The FPGA cross-correlates the digitised samples of every component from each channel, computes the Stokes parameters (polarization) – {\it I, Q, U} and {\it V} – and finally outputs these data parameters to a PC where they are time stamped with GPS timing precision to which the PC is synchronized by having a dedicated GPS receiver.
The digital backend solution follows the current trend, taking advantage of FPGAs flexibility capabilities, capable of providing the core of multi-purpose, re-programmable solutions to a wide variety of applications or just being building blocks of more complex signal processing systems. FPGAs are extensively used now in receiver backends, from Earth sensing radiometry digital backends (Bosch-Lluis et al, 2006; Camps et al., 2006) to radioastronomy applications for ALMA, APEX, MINT among others, like digital correlator projects from radio to millimetre waves like  (Fowler et al. 2005; Chang et al. 2005; Woods et al. 2009) beamformer boards and spectrometers (Herzen 1998; Hotan 2008; Klein 2006 among others). In particular, FPGAs are core to processing solutions prior to final centralized processing. We cite top performance efforts led by the Center for Astronomy Signal Processing and Electronics Research (CASPER) at Berkeley and Radionet UNIBOARD\footnote{http://www.radionet-eu.org/uniboard} that will impact the next generation of radio telescopes and their wide surveys like SKAMP\footnote{http://www.physics.usyd.edu.au/ioa/Main/}, MWA\footnote{http://www.haystack.mit.edu/ast/arrays/mwa/}, LOFAR\footnote{Http://www.lofar.org} and the future Square Kilometer Array\footnote{ttp://www.skatelescope.org}. The MINT (Millimeter INTerferometer) CMB instrument was probably one of the first attempst to use cost effective FPGAS for the core of digital backend based on design of the OVRO digital correlator. However, as far as we know from the literature, this constitutes the first application of a digital backend for polarimetry detection in the field using COTS components and a very competitive priced FPGA.

\section{Digital cross-Correlator}
\label{sec:2}

The requirements for implementing a full digital cross-correlator capable of computing the four stokes parameters would require a medium size FPGA with less then 40 multipliers and less than 5000 logic elements (LE) and we have no need for any of the special features available on the high performance FPGAs.  We selected an EP2C8Q208C7 FPGA from the Cyclone II family from Altera\footnote{http://www.altera.com} that at the time of design represented a trade-off decision between cost and performance. It contains 36 9-bit-multipliers, 8256 logic elements and we opted for a 208 pins QFP (quadrat flat pack) package.
The ADCs required in order to maintain a total complex signal bandwidth of 200MHz would have to digitize ideally at 200Ms/s, while in practice we have to allow a margin for the anti aliasing filter roll off. The ADCs selected along with the signal conditioning amplifiers implemented accept 0,5 V peak-to-peak signal, while the amplitude of signals we will use in operation will be about half of the ADC range. Since the best ADC performance occurs for differential signals, the signal conditioning before the ADC is a differential amplifier that transforms the single ended signal in a differential and provide the required common mode voltage to the ADCs. Digital signals going into the FPGA have already been interleaved inside the ADCs, so each ADC will present 16 bit to the FPGA that contains two samples at each FPGA reading cycle corresponding to sample n-1 and sample n. This allows a slower transfer rate to the FPGA being now at 100MHz clock. Total data transferred from the 4 ADCs is 6.4Gb/s.
After processing and integrating several samples the data rate is slow enough to be transferred to a computer. The computer is a small form factor PC104 module (AMD LX800 500MHz from Kontron1) with 256 Mb RAM and a 1Gb chip flash hard disk that interfaces the FPGA by the ISA bus. Although PCI bus is also available, we chose ISA bus communications for simplicity of processes. This PC is responsible to format the data, flag it and store it locally and make it available through the network via Ethernet IP using SFTP (secure file transfer protocol). The FPGA configuration was developed in VHDL language using Quartus II v6.1 web edition software. The prototype was implemented on a four layer PCB containing the ADCs and FPGA. High speed digital design considerations were applied during this design stage. The prototype board is presented in Figure 2.
\begin{figure}
 \includegraphics[width=1\textwidth]{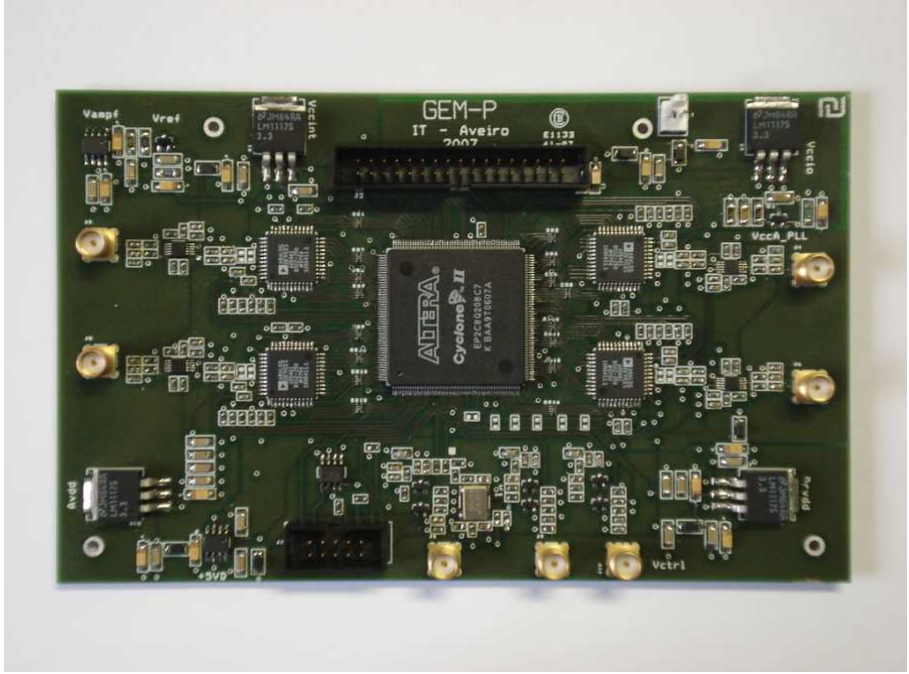}
\caption{Digital correlator board prototype PCB. One can distinguish easily the four ADC’s, symmetrically disposed around  the FPGA.}
\label{fig:2}       
\end{figure}
One of the main concerns during the PCB design was to keep all the signal paths distances equal in order to avoid phase differences between signals. This is why the design is symmetric in their layout appearance; there are four ADCs, equally separated on the right and left side of the board, corresponding to the left and right hand circular polarization, respectively. On the upper end is the ISA interface and on the lower end are the clock signals, triggers and Programming connector of the FPGA.
The PCB has four layers. We use the TOP layer mainly for signals, all the analog and digital signals. Immediately below is the ground layer, so all the lines can be considered microstrip transmission lines. The third layer is for power supplies. For the present design we have four different voltages being distributed, $V_{CCIO}=3$,$3V$, $V_{CCINT}=1,2V$, $V_{AMP}=5V$ and $V_{REF}=1V$.

\subsection{The calculation of the Stokes Parameters}

The four Stokes parameters {\it (S) I,Q,U,V}  describe the polarization state of any electromagnetic wave (Heiles et al., 2004 and references therein). The {\it I} parameter is the total power received by the antenna in the two polarizations, {\it Q} and {\it U} quantifies the power components of the linear polarization and {\it V} quantifies the amount of power in the circular polarization. Using the circular polarization base formalism, the parameters can be described below as:
\begin{eqnarray}
I &= &E_{LCP}E_{LCP}^*+E_{RCP}E_{RCP}^* \nonumber \\
Q &= &2Re(E_{RCP}E_{LCP}^*) \nonumber \\
U &= &2Im(E_{RCP}E_{LCP}^*) \nonumber \\
V &= &E_{LCP}E_{LCP}^*-E_{RCP}E_{RCP}^*
\end{eqnarray}

The EL and ER indicate the electric field of the left and right circular polarizations. These equations will be implemented as the algorithm core running in the FPGA for correlating the digitalized samples.

\subsection{FPGA Signal Flow Description}

The signals coming from the four ADCs ($ADC_1$, $ADC_2$, $ADC_3$ and $ADC_4$), enter the FPGA. Then a Stokes parameters calculation is performed followed by integration. After integration the data rate is considerably slower and then all data contained the polarimetric information is outputted to a PC104 [5]. A block diagram of the generic data flow from the ADC to the PC104 is shown in Figure 3.
\begin{figure}
 \includegraphics[width=1\textwidth]{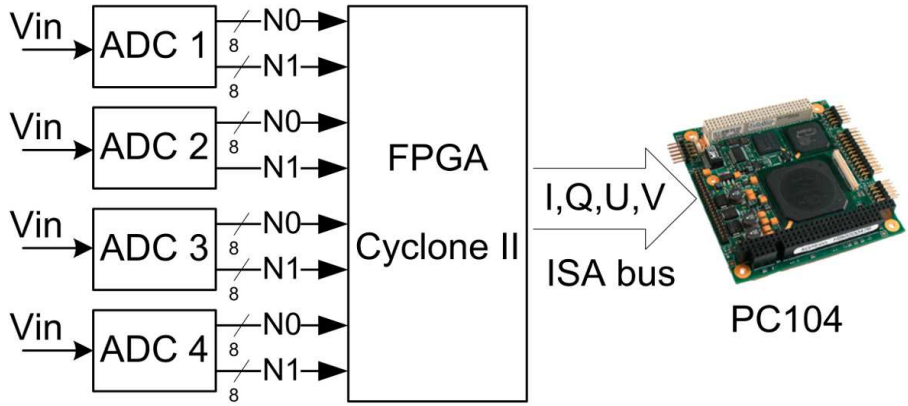}
\caption{Generic data flow and interconnection of the FPGA.}
\label{fig:3}       
\end{figure}
To fully represent a signal when it is converted to baseband (or to zero IF) a phase and quadrature detection for each polarimeter channel (RHCP and LHCP) is performed. These baseband signals can be considered a complex representation of the original signal by its real and imaginary parts. Therefore we have at the ADC input the signal components as detailed in table 1.

\begin{table}
  \begin{minipage}[t]{0.35\linewidth}
    \label{tab:1}       
    \begin{tabular}{ll}
      \hline\noalign{\smallskip}
      $ADC_1$ &  $Re(E_{LCP})$\\
      $ADC_2$ & $Im(E_{LCP})$\\
      $ADC_3$ & $Re(E_{RCP})$\\
      $ADC_4$ & $Im(E_{RCP})$\\
      \noalign{\smallskip}\hline
    \end{tabular}
  \end{minipage}

  \begin{minipage}[t]{0.65\linewidth}
    \caption{Components of the signals $E_{LCP}$ and $E_{RCP}$ entering the ADCs}
  \end{minipage}

\end{table}

Considering equation 1 and the information in Table 1, it is possible to define how the Stokes parameters are calculated from the signals in the four ADCs, equation 2.

\begin{eqnarray}
I &= & ADC_1^2+ADC_2^2+ADC_3^2+ADC_4^2 \nonumber \\
Q &= &2*(ADC_1*ADC_3+ADC_2*ADC_4) \nonumber \\
U &= &2*(ADC_1*ADC_4-ADC_2*ADC_3) \nonumber \\
V &= &ADC_1^2+ADC_2^2-ADC_3^2-ADC_4^2
\end{eqnarray}

Since the output of the ADCs present two samples N0 and N1 at each cycle, Stokes parameters would have to be calculated simultaneously for both N0 and N1 according to equation 3 and 4.
\begin{eqnarray}
I &= & ADC_{1N0}^2+ADC_{2N0}^2+ADC_{3N0}^2+ADC_{4N0}^2 \nonumber \\
Q &= &2*(ADC_{1N0}*ADC_{3N0}+ADC_{2N0}*ADC_{4N0}) \nonumber \\
U &= &2*(ADC_{1N0}*ADC_{4N0}-ADC_{2N0}*ADC_{3N0}) \nonumber \\
V &= &ADC_{1N0}^2+ADC_{2N0}^2-ADC_{3N0}^2-ADC_{4N0}^2
\end{eqnarray}

\begin{eqnarray}
I &= & ADC_{1N1}^2+ADC_{2N1}^2+ADC_{3N1}^2+ADC_{4N1}^2 \nonumber \\
Q &= &2*(ADC_{1N1}*ADC_{3N1}+ADC_{2N1}*ADC_{4N1}) \nonumber \\
U &= &2*(ADC_{1N1}*ADC_{4N1}-ADC_{2N1}*ADC_{3N1}) \nonumber \\
V &= &ADC_{1N1}^2+ADC_{2N1}^2-ADC_{3N1}^2-ADC_{4N1}^2
\end{eqnarray}

The stokes parameters would now require an integration over a long time to obtain the estimated pixel polarimetric information. This can generally be expressed as in the equation 5.

\begin{eqnarray}
\langle I\rangle &= &\langle ADC_{1N}^2+ADC_{2N}^2+ADC_{3N}^2+ADC_{4N}^2\rangle \nonumber \\
\langle Q\rangle &= &\langle 2*(ADC_{1N}*ADC_{3N}+ADC_{2N}*ADC_{4N})\rangle \nonumber \\
\langle U\rangle &= &\langle 2*(ADC_{1N}*ADC_{4N}-ADC_{2N}*ADC_{3N})\rangle \nonumber \\
\langle V\rangle &= &\langle ADC_{1N}^2+ADC_{2N}^2-ADC_{3N}^2-ADC_{4N}^2\rangle
\end{eqnarray}
\subsection{Algorithm Implementation}

The code was written in VHDL and used a behavioral description coding style. This made the overall design considerably easier to read and understand. However carefull attention to timing details were necessary.

As seen from the simulation, figure 4, the 8 bit data samples received by the FPGA are spaced in time by 10 nanoseconds and valid at the ascendant clock transition that should happen at the middle of the time where data is stable. If the data flow could be implemented as simple as depicted from the data flow previously described all the Stokes parameters must be calculated on each transition and produce a result every 10ns.
\begin{figure}
 \includegraphics[width=0.75\textwidth]{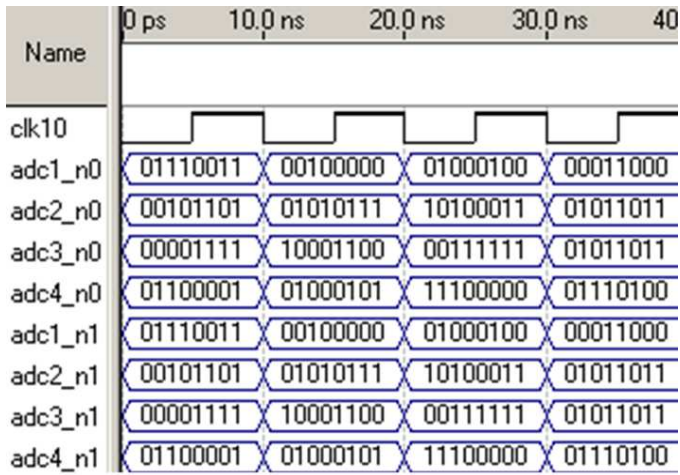}
\caption{Data coming from ADCs and entering the FPGA.}
\label{fig:4}       
\end{figure}

Figure 5 shows a block diagram of the FPGA implementation by representing it in four well defined data stages; the input registering of data and clock synchronization, the calculations of stokes parameters, the integration and finally the output data interface.
\begin{figure}
  \includegraphics[width=0.75\textwidth]{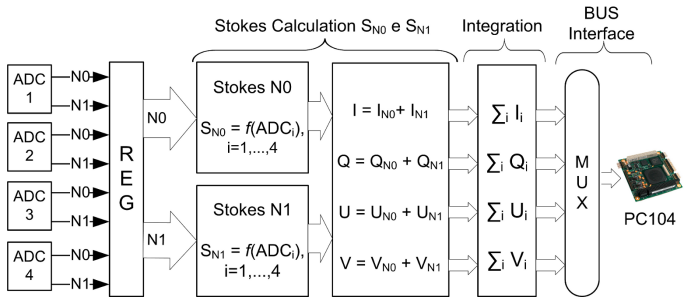}
\caption{Overall block diagram of the FPGA implementation.}
\label{fig:5}       
\end{figure}
\subsection{Input synchronization}

The very first task before performing any calculations consists on the synchronization of all the inputs, in order to guarantee that all data exists stable during a complete FPGA clock cycle, regardless of residual delays from the ADC-FPGA interface. Basically the data from the ADCs is registered by a skewed FPGA clock that is phase aligned with the center of data available from ADCs. A second register stores this data with the main FPGA clock. Now all data is available at FPGA clock and all samples now computed at the same time have the maximum time window available to be processed by the multiplication/addition logic necessary to implement the stokes parameter calculation.

\subsection{Stokes Parameters Calculation}

Since the FPGA receives two consecutive signal samples, the next step must implement a parallel processing of both samples. While the parallelization concept is easy to understand the simulation/verification was actually quite morose since multiplications/additions in general do take a predictable time to compute but they may compute wrongly if we do not allow enough time for signals to stabilize. This is a considerable problem to simulate/debug since different data values may take slightly different times and if we are on the edge of the computation timing, wrong computations may happen. As a demonstrating example, we show, in figure 5, that a square of a sample with eight bits needed 10.6ns to perform valid data when we would need less 10 ns for all the processing of equations in (3) or (4). As referred above, one can see that although the first result is correct the following calculations are incorrect.
\begin{figure}
  \includegraphics[width=0.75\textwidth]{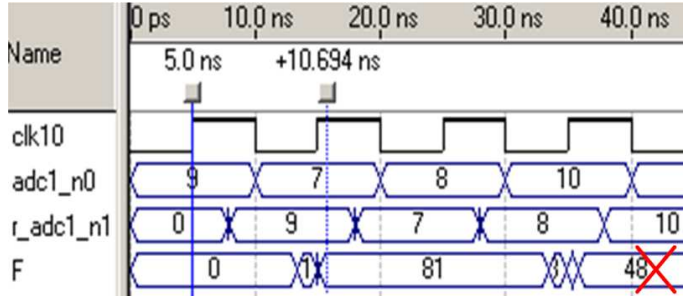}
\caption{Simulation showing squaring takes more than 10.7ns for correct calculations.}
\label{fig:6}       
\end{figure}

The simulation indicates we need 3 ns for the propagation time therefore only 7 ns will remain to perform the multiplications, which is not possible in a commercial FPGA with a speed grade of 7, as advertised by the manufacturer (Altera).

The solution is to slow down once more the speed of calculus by paralleling more functions in the FPGA (note that first parallelization occurred inside the ADC). Such parallelization of the data in two registers with twice more bits would extend the available time to 20ns Figure 8.

\begin{figure}
  \includegraphics[width=0.75\textwidth]{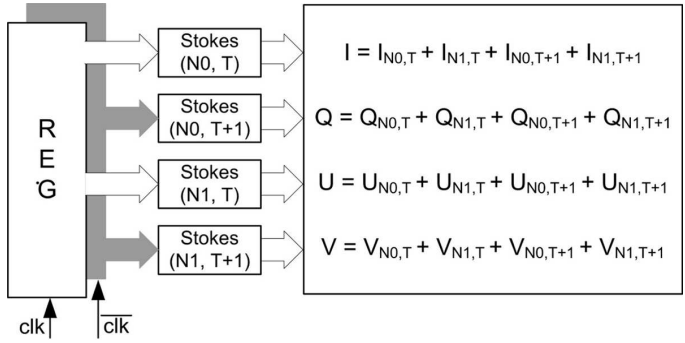}
\caption{Further interleave of data to extend the time to 20ns, note that registers are now clocked at half speed i.e. 50MHz.}
\label{fig:7}       
\end{figure}

To detail this process we shall consider here only one ADC. Assuming that ADC1 is in the N0 channel, all the samples are saved in the register {\it r\_adc1\_n0\_t1} on the clock ascendant transition and in the register {\it r\_adc1\_n0\_t2} on the descendant transition.  Figure 9 show the two registers now containing the information of one ADC of two samples during 20 ns that is half the initial clock speed.

\begin{figure}
  \includegraphics[width=0.75\textwidth]{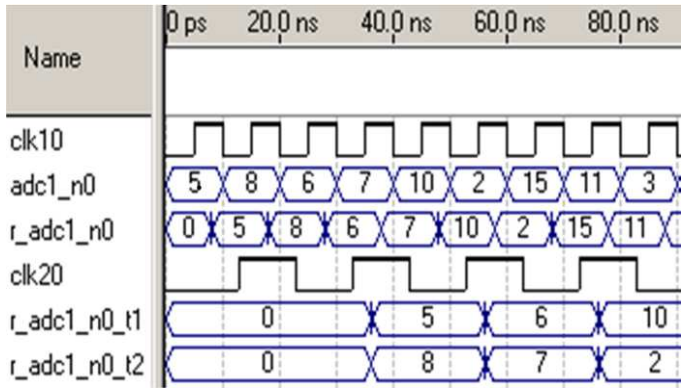}
\caption{Data extended to 20ns by paralleling the processing of more samples of each ADC.}
\label{fig:8}       
\end{figure}

The next step is to perform the calculations of each Stoke Parameter for channels N0 and N1 at the time instants T and T+1, corresponding to the two registers as in the following expression :

\begin{equation}
S=S_{n0}(t)+S_{n1}(t)+S_{n0}(t+1)+S_{n1}(t+1)
\end{equation}

\subsection{Stokes Parameters Integration}

The techniques employed above to allow the calculation of the stokes parameters have slow down the data rate considerably but increased the number of bits. The speeds required for the data transfer are still orders of magnitude slower than that to be used by the PC buses, ISA or PCI, available on the PC104. The goal is now to integrate the stokes parameters producing a statistical estimation of its value. This is accomplished by performing successive additions on each of the parameters to extract their average. In Figure 10 we show a diagram of the architecture used to obtain this data integration.

\begin{figure}
  \includegraphics[width=0.75\textwidth]{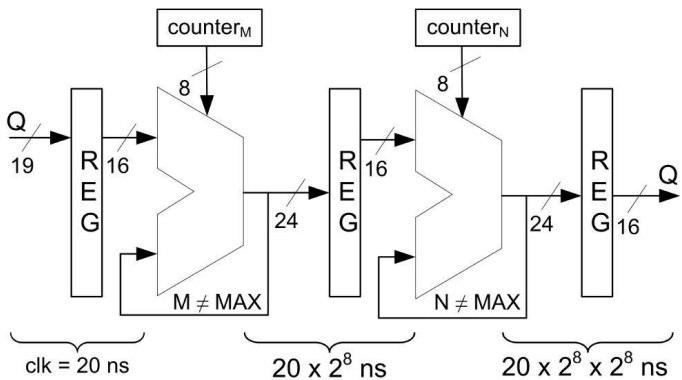}
\caption{Stokes parameters integration and data size reduction by cascading of two accumulators.}
\label{fig:9}       
\end{figure}

The processing undertaken by one parameter is shown in figure Figure 9. The others parameters will have the exact same structure. The input is the calculated stokes parameter and the output is a time integrated version, average, of the input. 
The input value is represented by 19 bits and here becomes reduced to 16bit by the first register that will only use the sixteen most significant bits (MSB) of this value. Apparently the accuracy is lowered, however we are processing a signal resulting from computation of 4 times 8bit signals that will not use the total dynamic range and do not require the full instantaneous resolution at any of the stages before integration. At early stages the signals are still random noise like signals and the fine resolution and sensitivity only appear after significant time integration. The first accumulator performs M sums of the parameter. This procedure is repeated again in a second stage to reduce the amount of information and at the same time to further integrate each parameter. At each integration level the time is increased by $2^8$, so in the final register we have the most significant 16 bits at a rate of $20 * 2^8 * 2^8$ = 1310720 ns ≈ 1.3 ms.

\subsection{Bus Interface}

With data available each 1.3 milliseconds, it is now possible to use one of the provided buses of the PC. For simplicity, we have chosen the ISA bus. The interface diagram is represented in figure 10. Interfacing the ISA bus using an 8 bits bus width, each parameter (that has 16 bit) will have to be transferred by two I/O bus cycles, one for the eight MSB and again for the least significant bits (LSB).

\begin{figure}
  \includegraphics[width=0.75\textwidth]{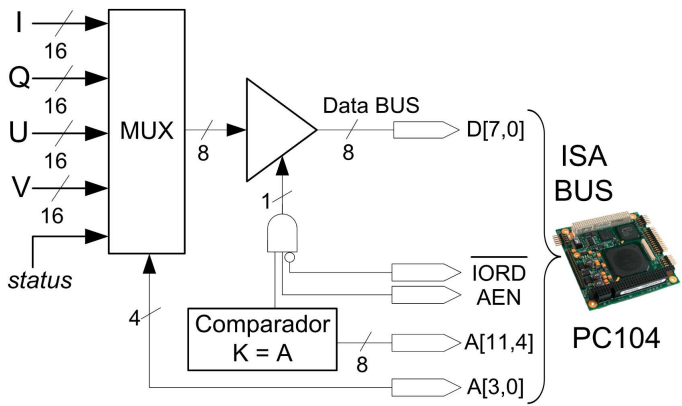}
\caption{ISA bus interface diagram.}
\label{fig:10}       
\end{figure}

To select all the parameters to be transferred on the bus read cycle we need 8 I/O address positions. A status register from the FPGA to the PC was also considered to retrieve the communication status and other internal flags. This action results in an address range exceeding 8 positions. Therefore 4bits of addressing were considered. The four LSB of the address bus are used by the multiplexer to select which parameter is sent to the bus. The control bus follows the classical ISA I/O signaling protocol, i.e. the Input/output read (/IORD) and the Address Enable bits. The selection of this PC peripheral (the FPGA as seen by the PC104) is made at a at an address range of the traditional I/O space that was found free.
The full FPGA implementation was test benched and calibrated, using progressively higher data rates starting at 24 MHz, 50 MHz 75 MHz and finally 100 MHz. 
The full code implementation of the FPGA used 16\% of the logic elements  and 89\% of the embedded multipliers. 

\subsection{PC104 Software }
The PC104 code was developed with basic I/O, time stamping and further processing of the samples. To efficiently control these tasks, a new micro-Linux distribution was generated in which high performance and low read/write memory needs were obtained so that a fan-less PC104 with solid-state disk could be used. It  avoids, as much as possible, hard disk writing and includes ramdisk (/var and /tmp $\rightarrow$ /dev/ram0), a deactivated swap area to avoid flash memory writings, since writing is a slow process on flash disks. It includes also interfaces for Ethernet and external micro-controller, C code for added correlator data acquisition and sample timestamping synchronized via a dedicated GPS receiver, code for interaction with the antenna attitude control system, data flagging and transport of other physical housekeeping data from the receiver.  No gcc compiler is included, since the chip is not meant for code development and compilation and makes use of Busybox 1.5 for further size-optimization. With only one ext3 partition and absence of  X system  it reduces the final distribution size to less than 58Mb.  This distribution is now called LIRAe for “Linux for Radio Astronomy embedded” and is based on a Red Hat Linux 8.0 with a 2.4x kernel and is actually very suitable to be used in any small embedded system requiring small PCs and extremely low disk access despite intensive data operations. Available upon request at http://www.av.it.pt/gem.

\section{Conclusion}

A new digital backend with a complex correlator implemented in a Cyclone II FPGA from Altera was developed for a polarimetry application on the GEM Portugal project. This implementation is suitable to be used in other sky mapping or radio astronomy applications or in any intermediate processing chain of the parallel processing of several signal channels. 
The time required for multiplications– along with the throughput required led us to use several interleaving mechanisms to slow down the speed enough to keep this design feasible on a low cost FPGA. The large time integration and subsequent slower data rate at the FPGA output allowed also the usage of a low cost PC104 solution and the usage of a traditional bus interface. 
This digital backend using competitive priced COTS technology forms a full complex correlator and it is suitable for large sky area surveys. 
To the best of our knowledge this is the first application of a full digital backend for polarimetry detection in the field.

\begin{acknowledgements}
The team acknowledges the Portuguese Foundation for Science and Technology (FCT) for the financial support through projects PTDC/CTE-AST/65925/2006, POCI/CTE-AST/57209/2004. RF acknowledges support from FCT through a PhD grant. DB is supported through a Ciência 2007 grant, funded by COMPETE and QREN programs. MB acknowledges support from PANORAMA project.
\end{acknowledgements}



\end{document}